# *A review and evaluation of the use of longitudinal approaches in business surveys*

Paul A. Smith*
S3RI and Dept of Social Statistics & Demography, University of Southampton, Highfield, Southampton, SO17 1BJ, UK
p.a.smith@soton.ac.uk

and

Wesley Yung
Statistics Canada, Tunney's Pasture, Ottawa, K1A 0T6, Canada
welsey.yung@canada.ca

**Abstract**

Business surveys are not generally considered to be longitudinal by design. However, the largest businesses are almost always included in each wave of recurrent surveys because they are essential for producing good estimates; and short-period business surveys frequently make use of rotating panel designs to improve the estimates of change by inducing sample overlaps between different periods. These design features mean that business surveys share some methodological challenges with longitudinal surveys. We review the longitudinal methods and approaches which can be used to improve the design and operation of business surveys, giving examples of their use. We also look in the other direction, considering the aspects of longitudinal analysis which have the potential to improve the accuracy, relevance and interpretation of business survey outputs.

**Introduction**

Business surveys (or more generally *establishment* surveys which can cover, for example, farms, and institutions such as schools and hospitals) are generally undertaken by National Statistical Offices (NSOs) (rather than other research organisations) for a number of reasons.

These include that they depend on register information from administrative systems to which NSOs, but generally not the public, have access, and that they often collect sensitive data, which is greatly aided by a statutory response status than can be mandated by NSOs. The population of businesses also has quite different properties to populations of individuals. Most noticeably the population size distribution is very skewed, with many small businesses and a few very large ones, with the large ones making up such a large proportion of activity that it is essential to gather information from them in order to produce high quality estimates. This has led to a suite of survey (and associated) methods constructed specifically for business surveys. Rivière (2002) reviewed the characteristics that differentiate business statistics from statistics about other types of populations, and there has been a sequence of conferences in the International Conference on Establishment Surveys (ICES) series where researchers and practitioners meet to discuss and present these tailored methods (Smith & Phipps 2014).

Business surveys are not generally considered as belonging in a longitudinal framework, although there are exceptions. For example the Australian *Business Longitudinal Survey* which ran for four years from 1994-5 to 1997-8 (ABS 1999), which was a true longitudinal survey of small and medium sized businesses with supplementary samples of new births each year, and which is still widely analysed (Hansell & Rafi 2018). A similar *Longitudinal Small Business Survey* was instituted in the UK in 2015 (BEIS 2018). Although business surveys are rarely designed as longitudinal surveys many of them exhibit characteristics that are longitudinal in nature. When considering business surveys it is helpful to separate them into sub-annual and annual (including those surveys which are less frequent than annually) surveys as they tend to serve different purposes. Sub-annual surveys, commonly monthly and quarterly, are typically undertaken to produce short term trends, while annual surveys are commonly more detailed and produce structural statistics on businesses. Given the objective of sub-annual surveys to produce trends, panel designs are commonly used while annual surveys tend to use independent samples with overlap control. These sample designs are implemented to fulfil specific objectives and while they are not designed to be longitudinal, they do exhibit some aspects of longitudinal surveys. In this paper we describe typical business survey designs and indicate areas where longitudinal approaches have been applied and note where their application could improve existing methods. Sections from "sample design and selection" to "time series analysis and relationships with longitudinal designs" cover the main elements of a business survey, broadly dealing with topics where longitudinal considerations are most relevant. In the "longitudinal analysis" section we change perspective and consider which elements of longitudinal analysis could be more widely applied to business survey data. The final section provides some discussion of the synergies of business and longitudinal surveys.

**Sample design and selection**

*Completely enumerated strata*

Populations of businesses are almost always highly skewed, with a small number of very large businesses accounting for a large proportion of the activity, and a large (sometimes very large) number of small businesses accounting for a relatively small proportion of activity (Rivière 2002, Smith 2013 Table 5.1). Hence, they are typically stratified by size with the largest businesses placed in a completely enumerated stratum. Given the importance of these large businesses to the overall estimates, even in rotating panel surveys (see below) these largest businesses are included every period (for a counterexample see Zimmermann *et al.* 2018). Thus, even though the goal has not been to design a longitudinal survey, business surveys do contain longitudinal samples of the largest businesses as a consequence of the requirements of users for the highest quality statistical outputs, and the way these affect the design process. Such longitudinal data is not a representative sample, but acts like a cut-off sample, covering much of the business activity.

*Rotating panel designs*

Sub-annual business surveys commonly follow a fixed sample of units in order to effectively estimate trends with a sample of births added to improve the cross sectional estimate. In addition, some surveys include a rotation component to reduce respondent burden on the small and medium businesses. This design resembles a rotating panel survey with a single panel being used at any one time – that is, businesses enter and leave the panel at each wave (survey occasion), but do not form a coherent panel which will necessarily remain in the survey for the same period. This is in contrast to the rotating panel designs used in some social surveys, which use multiple panels, each following a fixed pattern of occasions in the survey (Smith *et al.* 2009; see also the "coordinated sampling" section below). The rotation of units can then be induced in different ways:

(a) by selecting a new panel with conditions on the overlap with the preceding panel (as in the US Consumer Price Index collection, Johnson *et al*. 2012), which suggests a fixed panel for some period, with infrequent changes because of the effort involved;

(b) by having regular rotation of units in and out each period so that the sample size is maintained, often based on a coordinated sampling system (Ohlsson 1994, Lindblom 2003).

There is a trade-off between the length of stay in the panel and the resources required to collect data. Longer stays in the panel increase the chance of non-response from businesses, but give fewer new businesses to the sample each period, which means lower costs for contacting respondents to gain their initial cooperation and provide a certain amount of help or training in responding to the questionnaire. An example of a change in response to this issue comes from the Office for National Statistics, which changed the rotation period for most businesses in its monthly surveys from 15 months to 27 months in the mid-1990s, largely to reduce the costs associated with businesses added to the sample. For business surveys at Statistics Canada, method a) is typically used. Monthly business surveys usually follow a fixed panel of units, with a sample of births each month, for a period of four to five years at which time the panel is renewed with a controlled overlap to ensure that old and new samples contain some common units other than those in the completely enumerated strata (see "completely enumerated strata" above). See also "coordinated sampling" below for methods used to control this overlap. On the other hand, annual surveys are normally renewed each year, again with a controlled overlap.

Social survey researchers have given considerable attention to rotation group bias (RGB; Bailar 1975) – the property that a unit's response to a question depends on how many times it has been included in the survey. In principle this effect might be expected in business surveys, and Srinath (1987) and Sigman & Monsour (1995) both mention it as an issue to be considered. However, we have not located any published examples where an attempt has been made to estimate RGB in a business survey. The general feeling with short term business surveys is that, since typically a small number of questions is asked (less than five), there is not a significant rotation group bias other than potentially a 'burn-in' bias when new units are introduced into the sample. However, to combat this burn-in bias in sub-annual surveys, parallel runs are common where new units are surveyed for several periods before their data is assessed to be satisfactory and can be used in estimates, allowing the original units to be dropped. These restratifications where information used to stratify businesses is updated and the sample, or panel, is renewed (again with a controlled overlap with the old sample) occur only every four to five years.

An exception to the single panel approach was found in the US Monthly Retail Trade Survey (MRTS) and Monthly Wholesale Trade Survey (MWTS), which both used a multiple panel design for smaller businesses (and a completely enumerated stratum for larger ones) from before 1963 (Woodruff 1963) to 1996 (Cantwell & Caldwell 1998). Originally the selected sample was divided into 12 panels, one of which was surveyed each month so that a business responded to questions once per year, but later this changed to a division of the selected sample into three panels with each business answering questions quarterly. In both cases information on turnover was collected for *two* months, the current and the preceding one. This led to two different biases, neither of which is (classical) RGB. First, the panels were retained for five years between economic censuses, and this meant that they changed, through births and deaths and through growth and contraction of businesses, so that it was possible for panels to become unbalanced, and this induced a periodicity into the estimates. Second, in collecting two months of data at the same time, the previous month's turnover data was typically based on accounting information and the current month's turnover data was more provisional. The current month's turnover was on average always lower – a systematic measurement error, which induced large revisions. These surveys were therefore restructured to a fixed one-panel pure monthly survey to avoid these challenges (Cantwell & Caldwell 1998).

In the 1990's Statistics Canada did use a multi-panel design in their Agriculture Crops surveys. Panels were created in such a fashion that multiple panels could be used for different occasions of the series of Crops surveys. By choosing different panels, sample overlap and rotation could be controlled. For more information on the sample design, see Julien and Brisebois (1999).

***Coordinated sampling***

Several characteristics of business surveys make the approach to longitudinal sampling different from social surveys. First, there is generally a good quality register of businesses, so the list of businesses in the population is quite well-known, and the register information can be used in designing samples – indeed it is important because of the variation in business size.

Second, businesses change relatively more rapidly than human populations – for example in the early 2000s, only half of European businesses survived five years (Schrör 2009) – and businesses may grow or shrink rapidly, which means that they can move between strata in a stratified design. Thirdly, larger businesses are typically included in multiple surveys, and there has been a desire to control the survey burden – the cost to the business of responding to the surveys.

Business populations in some industrial classification by size strata are small, and panel inclusion tends to be long compared to social survey panels (27-120 months in ONS business surveys and 48 to 60 months for Statistics Canada surveys). Attempting to have a representative panel added to the survey on each survey occasion, would require a sample design with a substantial sample size in each stratum (at least 27 using the above examples) just to have a single business from that stratum in each panel. So it is generally unrealistic to have a series of panels each with the same design. Instead a single panel is used in each stratum, with rules to govern how this panel changes. One approach to manage this is to assign a permanent random number (PRN) to each business when it joins the sampling frame, and then to sample a range of PRNs, with the range "moving" incrementally to allow the sample to change composition gradually. The rate of movement governs the overlap (positive coordination) – slower movement means greater overlap. One way to get negative coordination (no overlap) between surveys is to restrict them to different PRN ranges. There are many other approaches to coordinated sampling (see for example Nedyalkova *et al*. 2006), and this seems to be an active area of research with a range of different ideas and mechanisms still being proposed and implemented (see for example Smeets & Boonstra 2018, Gros & Le Gleut 2018).

The use of these methods of sample coordination can in some cases induce a particular design in the longitudinal sample (see Nedyalkova et al. 2009; for example some PRN-based designs produce Poisson samples for the cross-sectional design and induce systematic samples in a longitudinal design), but the constant changes to the stratification also contribute to make the process less balanced than in multiple panel designs. But the major advantage of these approaches is that each cross-sectional design is well-controlled (that is, the cross-sectional sample is a random selection with the required cross-sectional design features), and this is typically the objective of a business survey as it allows for the calculation of high quality cross-sectional estimates.

Coordinated sampling is strongly connected with business surveys, but it is also used in social surveys in some cases, for example in the Swiss Federal Statistical Office. It is one area where there is potential to borrow methods from business surveys to extend the possibilities for longitudinal social surveys, where it could be used in selecting top-up samples, or in setting up rotating panel type designs.

*Longitudinal samples*

The net result of the sampling procedures described in 'completely enumerated strata' above is that many business surveys have longitudinally linked sample units, which can be used to explore and analyse transitions and developments in businesses and their activity. But the characteristics of these samples are somewhat different from those usually met in longitudinal analysis, so that additional thought must be given to issues of representativity (when only the larger units are linked), the way births and deaths appear in the samples, and the constant change induced by coordinated sampling.

The features described in this section mean that longitudinal design considerations in social surveys, which are typically focussed on clustering for fieldwork efficiency, oversampling specific populations and ensuring adequate initial sample sizes to protect against differential dropout (Smith *et al*. 2009), have a quite different focus.

*Recruitment and maintenance*

The recruitment of sample members is generally quite different, since in business surveys participation is generally compulsory, though they still suffer from nonresponse. Particular efforts are made to obtain responses from the largest businesses since they contribute disproportionately to the estimated totals; this usually leads either to a response or enough information to make a good imputation, so a longitudinal dataset formed from completely enumerated strata may be almost complete even when the overall survey response rate indicates some nonresponse. Dedicated personnel or teams may be involved with the largest and most complex businesses (for example Sear *et al*. 2008), and this can forge strong communications between the responsible individuals in the surveying and respondent organisations in the same way that longitudinal social surveys can (possibly) benefit from having the same interviewer contact the respondent on each occasion (Lynn *et al*. 2014).

Sample maintenance in social surveys often involves a keep in touch (KIT) exercise if there are long periods between interviews, but business surveys obtain contact information through administrative records, which mean that in principle this is not needed. There have been attempts to motivate businesses to participate in surveys by providing feedback on the results of the surveys, with mixed success (Snijkers & Jones 2013, p416; see also Holmberg 2004).

**Measurement processes**

Once the sample has been designed and recruited, the next stage is to collect information from the sampled units. This involves asking questions, which may be done using a number of modes, and the manner in which this is done may affect the quality of the responses.

*Questionnaire design*

There are some important principles in designing questionnaires for longitudinal studies, many of which apply equally in business and social surveys, for example keeping the questions and question order fixed, so that they elicit consistent information from respondents on each occasion. The type of information collected in business surveys tends to have a more objective, often numerical, definition, and therefore should be less affected by question order effects and seam bias (see "the response process and measurement errors" below) than some topics in social surveys. In addition, the information needs for business surveys tend to remain very stable over time which leads to very little change in questionnaire content.

There is also a risk to longitudinal continuity from changes in survey mode (Dillman 2009), which applies to both social and business surveys. Testing questionnaires in all the modes available to respondents is an important principle (Bavdaž et al. 2019,) and is designed to minimise and understand the impacts of such changes. Since regular business surveys have not generally been viewed as longitudinal, this process has not been viewed in the same way, but it is still necessary to deal with the effect of changing questionnaires in making estimates of period to period change (eg Bollineni-Balabay et al. 2016).

*The response process and measurement errors*

Probably the biggest difference in the response process is that many business surveys are compulsory, but this does not change the need for the questionnaires and other data collection processes to be well-designed and tested in order to elicit the required information. Seam bias, where changes in variables between waves are overstated and changes measured within waves are understated (Callegaro 2008) seems less likely for business surveys, where data can be collected immediately after the period to which they relate and typically refer to more objective measures, though it has been demonstrated in quantitative variables in a social survey by Conrad *et al.* (2009). Similar issues do arise in business surveys, however. For example in the UK Producer Price Index respondents are requested to provide the average price for an item in the reference month, but examination of the data indicates that few respondents make this calculation, and therefore the price changes all appear at the time the questionnaire is sent; in practice this is unlikely to have a significant impact on uses of a price index derived from these data, however. But a similar situation in an annual data collection might have an unacceptable effect. The example of MRTS and MWTS from "rotating panel designs" above could also be construed as an example of seam bias.

There is mixed evidence over whether proxy reporting increases changes (Callegaro 2008, section 4.1.1), but the same phenomenon could be better or worse in business surveys. A change in respondent may have minimal impact when respondents have detailed 'desk instructions' on how to calculate the inputs to repeating questionnaires (Bavdaž et al. 2019, section 3.3), which is usually the case with business surveys. On the other hand, where there are no such processes, or where questions are more subjective, for example in business tendency surveys (also known as business conditions or business confidence surveys), the change of respondent may cause a large and false measurement of change between survey occasions. Friberg (1992) and Davies & Smith (2001, section 6.4.1) give an example where a change of respondent effectively produced a reinterview study which demonstrated this measurement error.

Dependent interviewing is one strategy for improving estimates of change by reducing measurement errors (Jäckle 2009). It has not been common as a direct strategy in business surveys (though see Holmberg (2004) for a counterexample), because data from previous collections are sensitive and the perceived risk of sending this information openly is too great.

However, past data have been used regularly in data editing (see below), which could be regarded as reactive dependent interviewing. Where data collection is moving to electronic modes, having past responses stored securely so that they can be used in dependent interviewing processes becomes much more practical.

**Nonresponse, editing and imputation**

*Nonresponse*

The classical pattern of response in longitudinal surveys is of continuous response followed by attrition, although there are possibilities to re-engage non-respondents from previous waves. Patterns of nonresponse in business surveys, however, are much more likely to be "swiss cheese" with holes (non-responses) framed by some responses, because the focus on cross-sectional estimation means that a response is always valuable, and the statutory requirement to respond makes it less likely that a business will drop out of the survey entirely. Dealing with non-response is therefore more about imputation than about complex weighting methods, although we return to weighting in "Estimation" below.

Even the estimation of response rates may be done differently, since in business surveys a weighted response rate accounting for differences in the sizes of businesses is often more informative about the quality of the estimates, whereas in social surveys each person counts approximately the same.

*Edit and imputation strategies*

Editing in business surveys makes strong use of the time series of observations, and indeed this is the main focus in the classic paper by Hidiroglou & Berthelot (1986), although cross-sectional edits are also important. So clearly past values are important in defining unusual observations.

Business surveys make wide use of imputation, but this is generally with the aim of making the best possible cross-sectional estimates (although always with an eye towards the impact on movements). For this purpose imputation can proceed period by period, and there is no particular requirement for imputed values to be consistent from period to period, or consistent with actual responses (although consistency with previous responses often improves the cross-sectional imputations, see below). Longitudinal analysis however has a different focus where the consistency across time is important in order to estimate transitions and survival as well as possible. For these purposes it would be good to jointly impute values in multiple periods, in order to obtain the right joint distributions, and since we are then interested in preserving distributions, rather than just means and totals, a stochastic imputation is also needed. Stochastic imputation is rare in business surveys, but Smith & Perry (2001) describe one such example in a survey of business demography in central European countries. Such joint imputation means revisiting earlier imputations as more information becomes available, and this creates revisions, which are generally the bane of users of business statistics (Bean 2016, p21-22) and therefore undesirable, though this opinion might be moderated in a longitudinal analysis.

Weighting solutions for unit nonresponse are also available, but while these are relatively straightforward for cross-sectional estimation, defining appropriate calibration constraints is challenging longitudinally (see "estimation" below) – as it is in the longitudinal social survey context too (see Lynn & Watson (2020, forthcoming)).

*Models for longitudinal imputation*

Compared to longitudinal social surveys, longitudinal business survey data contain more auxiliary or prior information (from the business register or previous occasions of the survey), but the number of variables collected at each wave is much less (especially for sub-annual surveys). For imputation in cross-sectional business surveys this leads to ratio or regression type imputation models (Kovar & Whitridge 1995) based on the relationship between responses and auxiliary data for responding units, which is applied to the auxiliary data for the non-responders. In the rotating panels typical of short-period business surveys there is often some previous information from the *same* business, and it is generally believed that such information will be a better predictor for the missing value. Ratio imputations which use this previous information are common in business surveys (eg Services Annual Survey in the US (Thompson & Washington 2013)), and this type of imputation is called *historical imputation* by Kovar & Whitridge, although we could as easily consider it to be *longitudinal* imputation, since it follows a single unit over time using common trend information. This approach will be most effective where the relationship over time is stronger than the relationship between units, and the wide range of applications suggests that this generally true, although we have not found any published examples where this is actually tested.

Internal research at Statistics Canada showed that for particular surveys, ratio imputation involving a unit's previous value and an auxiliary variable works well.

The modelling carried out by ONS and described in Elliott (2009) gave some initial indications that deterministic regression imputation (which includes ratio imputation as a special case) based on the previous value was consistently better for imputation (ONS, unpublished report), but the models are not sufficiently detailed for this conclusion to be robustly demonstrated, and further research on this topic would be useful.

There aren't always previous values, so imputation can become a mixture of approaches (known as *composite imputation*), such as historical imputation for some cases, and ratio imputation to auxiliary variables otherwise (eg Shao & Steel 1999, Statistics New Zealand 2013). The mixture may be extended further if there are different patterns of missing values, which may require looking back several periods. Composite imputation also makes variance estimation in the presence of imputed values more difficult, and Shao & Steel (1999) set out an approach to this problem.

This naturally creates some further challenges for a strictly longitudinal analysis, because the imputation depends on the pattern of missingness, which may affect some of the types of variables of interest – for example whether businesses just starting up have different characteristics from established businesses. Some of this difficulty may be avoided by revising the imputations for previous periods (backward imputation, Hidiroglou & Berthelot 1986) once a response is received, which means that the imputations will be based on the correlations within a single unit's responses over time. However, this would lead to revisions which, as previously noted, are not appreciated by data users.

For a true longitudinal analysis we are also concerned with the distributional properties of changes – so imputing with a ratio imputation using the average (or some robust version such as the median) change will reduce the variance of the change and lead to poor longitudinal inferences. A stochastic imputation approach will be needed to create a dataset which has appropriate characteristics for estimating such transitions.

# Estimation

## *Estimation using common units*

The use of rotating panel designs and completely enumerated strata result in relatively many common sample units between consecutive survey occasions. The (sampling) covariance from having these common units can be exploited to reduce the variance of estimates of change in totals. One way to do this is to use only the common units in estimating the movements, which are then projected forward from a fixed point, usually provided by a benchmark survey with a larger coverage. The common units have been known informally as "matched pairs" because they have a pair of observations on a matched unit. The matched pairs approach has been used in official business surveys (Kokic & Jones 1998, Smith *et al*. 2003), but the units which do not have matched observations may behave differently, and therefore the estimates of movements may be biased. The series may therefore drift and needs periodic benchmarking. If the benchmark is available relatively slowly (for example from an annual survey), then the error from the drift since the most recent benchmark can outweigh the variance gain from using the matched pairs (Kokic & Jones 1998, Smith *et al*. 2003). By definition units that start or cease to trade cannot have matched observations, so matched pairs approaches miss this element of change in the business population.

A similar approach has been used in the US Bureau of Labor Statistics, where a *link relative estimator* estimates the movement each period and is projected forward from a benchmark (Copeland & Valliant 2007). This estimator has the same challenges around possible bias from units that do not match between periods, particularly births and deaths, and models are used to adjust for this (Mueller 2006). The risk of bias increases as distance from the benchmark values increases, arguing for frequent benchmarking (Copeland & Valliant 2007).

Composite estimation, where the matched and non-matched portions of the sample are weighted to give the estimate of the movement with the smallest variance, has also been used occasionally in business surveys. For example it was used in the US Monthly Retail Trade Survey and Monthly Wholesale Trade Survey when those had multiple panel designs as described in "rotating panel designs" above (Cantwell & Caldwell 1998). It does not seem to have been widely used in business surveys more recently, though Preston (2015) has extended composite estimation methods to account for the more rapid change seen in populations of businesses.

The challenge with all of these estimation approaches which use elements of movement estimation is that the focus on estimating the change as accurately as possible causes a disconnection between the (sequence of) change estimates and the cross-sectional estimate of level (which, according to the design, is generally unbiased). For example, the composite estimate of the change is not the same as the difference between two consecutive cross-sectional estimates. This also has the potential to cause drift, and require the periodic benchmarking. An extra challenge of these methods is that the population is not stable, so that some account must be taken of the evolution of the population between benchmarks. However, with the modern emphasis on the use of administrative data in official statistics, some reconsideration of whether these approaches can be made operational by benchmarking to administrative data may be useful, and such an approach is considered by Mance (2016).

An alternative approach is to use a *regression composite estimator*, making the most efficient use of the matched and unmatched portions of a rotating sample using a combination of composite estimation and weighting (Singh, Kennedy & Wu 2001, Fuller & Rao 2001). This approach has been used in the Canadian Labour Force Survey, but we have not found any examples of its application to business surveys; although it was investigated for Statistics Canada's Survey of Employment, Payrolls and Hours, ultimately it was not implemented.

For sub-annual business surveys the user requirement has generally been more for a measure of change than for a measure of the precise level (Rivière 2002, p155, Smith *et al*. 2003 section 3.7). Therefore the mismatch may not matter too much. But a further consequence of occasional benchmarking is that the series is revised when it is benchmarked, and users of economic statistics generally dislike revisions. The combination has led to the replacement of matched pairs and composite estimates by repeated cross-sectional estimates in many cases. But for a specifically longitudinal analysis, there would be a case to revisit this approach, and to use it as a basis for describing the evolution of the estimates of the target variables.

*Changes in business structures*

Businesses are not organised according to any standard model, but in a way that suits each individual organisation. Administrative processes required by governments impose some constraints on this, and the use of administrative data is widespread in the construction of business registers. Therefore we need a *units model* to arrange these administrative data into a suitable structure for surveys (Colledge 1995). For any particular statistical collection a choice of which unit(s) to use has to be made so that the inputs and outputs can be as easy and relevant as possible. The choice of unit for any particular survey or analysis may well affect the estimates, and this has become known as "the unit problem" (van Delden *et al*. 2018, Lorenc *et al*. 2018).

According to the application of any particular units model, a 'business' does not necessarily consist of the same set of units between two periods. It may be affected by mergers or splits/demergers, or by acquisition or disposal of units. In the usual process of estimation for business surveys, these changes do not present particular challenges, as each period is considered separately, leading to cross-sectional estimates which use the structures in place in each period. The time series of such estimates may show spurious changes, resulting from the interaction of these structure changes with the classification rules in the units model – units may change classification according to which business they are part of. How much these changes are allowed to affect estimates can be controlled by the process by which the sampling frame is updated. In the UK a "frozen" register is used for monthly and quarterly surveys, with these structural and classification changes taken on once per year, with extra analysis of the impacts due to structure changes to separate it from the normal evolution of the series.

A longitudinal dataset constructed from a business survey is quite a different concept than a panel or cohort study, as it will not show the effects of *following rules* (rules about whether people are retained in the sample when they change households or in this case business units when they change ownership), which are designed to follow the units originally sampled. As far as we know a panel or cohort survey of businesses (or business units) has never been attempted in this way (although this might make for a very interesting dataset, showing the characteristics of business units being bought/sold and started up/shut down, and relationships with other units in the same business at the same time. It would also present some practical analysis challenges leading to further development of procedures). Therefore the need for weight share methods (Lavallée 2007, Lynn & Watson 2020, forthcoming), which allocate part of the weight of original sample members to new sample members which have been added because of the following rules in the same household/organisation etc, is reduced. The other bane of panel and cohort studies, estimating the eligibility of nonrespondents, is also much less, since the business register provides the information (though some of it with a lag, because of delays in reporting or updating. But if we are making a longitudinal dataset many of these lags will have worked out during its construction). Then we have the problem of varying inclusion probabilities, which becomes a calibration problem; the similarity of sampling in adjacent periods may mean that some averaging of selection probabilities followed by appropriate calibration to known population totals from the business register generates a satisfactory solution, but again there is little existing work in this area. Some similar questions can be addressed using post hoc longitudinal datasets, as described in "business demography" below.

By contrast, a long-term longitudinal database, or one incorporating information from more than one survey might be seen as a form of indirect sampling and therefore benefit from weight share approaches.

*Outlier detection*

A challenge which is more prevalent in business surveys than social surveys is that the highly skewed distributions of variables of interest mean that some observations have a disproportionate effect on estimates (whether cross-sectional or longitudinal); these effects can arise from the observation of a variable, but equally may be driven by the weight, or a combination of observation and weight. In business surveys these outliers are generally treated in some way to reduce their impact on estimates (eg Mulry *et al*. 2014, Clark *et al*. 2017). But outliers are often more obvious when considered longitudinally rather than cross-sectionally, as they can be compared in a time series.

*Variance estimation*

Variance estimation is a complex topic, and it suffices here to note that sampling in business surveys is almost never with negligible sampling fractions or ignorable, and that therefore many of the standard approximations for variances used in social surveys do not work (for example approximating the variance by the with-replacement sampling variance). For longitudinal analyses, these challenges add an extra layer of complication in inference for business surveys. For example, see Knottnerus & van Delden (2012) for variance estimators for overlapping business survey samples.

**Time series analysis and relationships with longitudinal designs**

*Seasonal adjustment*

Seasonality in longitudinal studies is a phenomenon that is less often assessed; when waves last at least a year, there is a relatively small amount of information for assessing within-year changes, though there are longitudinal social surveys with short period waves. Business surveys frequently have 'waves' at monthly or quarterly intervals, and the seasonality of the estimates from these is a key component of understanding the evolution of the measured phenomena. The theory and literature of seasonal adjustment is very large, and here we merely note its usefulness in interpreting longitudinal phenomena that are measured over periods less than a year.

*State space models for evolving phenomena*

The multiple panel structure of those few household surveys designed as rotating panels lends itself to modelling using state space models, which can model an underlying evolution and account for the rotation group bias (eg Pfeffermann 1991). Business surveys do not generally use multiple panels, however (see "sample design and selection" above), and therefore less attention has been paid to these types of model structures. Nevertheless there are possibilities to use these approaches, for estimation and adjustment of rotation group bias, and this would seem to be a fruitful area for further research.

One specialised type of business survey which we have not discussed so far are price surveys, the basis for the construction of price indices of different types, which involve the regular collection of prices from business sources (both for consumer and producer prices). Since it is not exactly the same copy of an item which is priced each time (the first one will probably have been sold by the second price collection occasion) this is not strictly longitudinal in the sense of following the same units through time, but can be regarded as pseudo-longitudinal, following an item through time. Dorfman (1998) introduces this approach and uses it as the basis for a state space model of the evolution of prices.

*Longitudinal analysis*

There is a large user requirement for information which can be used for the longitudinal analysis of business populations, from researchers and policymakers seeking to understand and influence business formation and survival to business experts gathering evidence on the success of businesses using certain management methods and strategies. The general lack of designed longitudinal surveys of businesses has led to the construction of analytical databases from other information. One of the main distinctions of business surveys is the existence of a good (statistical) business register derived from administrative sources and used as the basis for statistical collections. Business registers typically contain rather few variables of direct analytical interest, but because of their regular updating and good coverage of the population they provide a good, basic data source. They also provide a 'spine' to which other information can be linked to provide a wider set of analysis variables, though not in general covering the whole population.

Wider provision of analysis databases has therefore generally taken advantage of the repeated sampling of businesses in rotating panel surveys (see "rotating panel designs" above) and matched together the various responses to construct databases which cover longer periods and allow these sorts of longitudinal analyses to be undertaken. These databases could contain data coming from administrative sources as well as surveys. Databases of matched business data are available, usually under special license conditions because of the disclosive nature of business data, in many countries, eg Canada (Statistics Canada, 2016), New Zealand (Fabling & Sanderson 2016), Australia (Hansell & Rafi 2018).

There are substantial similarities in the types of analysis methods used in longitudinal business and social surveys, with estimation focused on gross and net changes between different times, and how these changes are related to other measured characteristics. Both need to use weights in the analysis to compensate for differences in sample representation, but these differences tend to be much greater in business surveys. Business surveys typically have a high proportion of numeric variables, which makes some analysis more straightforward, but also has skewed error distributions which complicates other analysis.

In the subsections below we change perspective, and instead of considering where methods designed for business surveys show longitudinal characteristics, we consider where methods of longitudinal analysis developed (largely) for social surveys could be applied to business data.

***Business demography***

There are plenty of examples of versions of the business register being used as the basis for longitudinal analyses (for example Ahn 2001, Bartelsman *et al*. 2005), though these rely on the assumption that the register is a reasonable reflection of the state of the business population. Lags in registration and deregistration processes, and the possibility that businesses may be dormant for a period make the validity of this assumption questionable in some situations. Bartelsman *et al*. (2005) discuss the challenges in producing reasonably consistent cross-national datasets because of the variety of definitions and rules associated with the business registers (and some other sources). But this can equally be a challenge within countries when the rules associated with registrations or the administration of the business register change. The Demography of Small and Medium Enterprises project (Smith & Perry 2001) is an example of where these issues were largely overcome. It provided an analysis of the survival and characteristics of business populations in twelve central European countries in the years following the breakup of the Warsaw pact, which involved a longitudinal survey based on the business register and had as its targets to estimate the ghost rates (proportion of businesses present on the register but not operational) and survival of businesses with different characteristics.

Of course the business register is not the only potential source of complete linked data on the business population, and there are examples of data from administrative sources being linked and made available for analysis, such as the Longitudinal Business Database in the US (Pivetz *et al*. 2001), which is derived from quarterly unemployment insurance microdata.

### *Gross flows*

Where data are linked longitudinally, it opens up the possibility to estimate the sizes of flows (transitions) between one state and another – for example business size in different categories. These statistics are readily calculated from longitudinal datasets once the weighting issues discussed in "edit and imputation strategies" are resolved. Gross flows have been more usually utilised to analyse employment change, but the estimated flows can provide substantial insights which are not available from the net statistics in other cases too such as how businesses grow and shrink, or how they shift between industrial classifications.

### *Survival analysis/Life course analysis/event history analysis*

Different transitions and events in the life of a business can be analysed by modelling with a longitudinal dataset, using suitable indicator variables to identify the states to be modelled.

These approaches have been used for businesses (eg Hill *et al*. 1996 use event history analysis with data on businesses taken from the stock market in the US).

The time to a transition (or the survival rates) can be related to different characteristics through survival analysis (eg Knaup & Piazza 2007), and this type of approach can be extended to Cox proportional hazards models and the Kaplan-Meier estimator.

These types of analysis generally require the use of a multilevel modelling approach to account for the correlation between occasions in the measurements from the same business. The common assumption of multilevel models that errors are normally distributed is unlikely to be good for most business populations, as the skewness of the business population feeds through into a skewed distribution of errors. Some data transformation or some more advanced approaches using different error distributions might therefore need to be investigated or developed to make these approaches really effective.

***Borrowing from business survey analysis to longitudinal surveys***

"Time series analysis and relationships with longitudinal designs" above showed some ways in which time series approaches can be integrated into longitudinal analysis, and this is one area which social surveys analyses have generally used sparingly, except where econometric approaches have been employed. In particular it may be interesting to relate longitudinal analyses to externally measured variables, to examine the impact of changing conditions on behaviours and transitions.

Businesses are also connected strongly through networks of transactions. These have traditionally been summarised in the national accounts, a framework showing flows of goods and services and money between different actors in the economy (for an overview see Bos (2013)). Although they are called "accounts", the use of survey estimates for many components means that the components which should balance (such as income and expenditure in accounting) rarely do, and this has led to procedures for balancing – finding the best adjustments to estimates accounting for their accuracy to give consistent accounts (eg Stone *et al*. 1942, Daalmans 2018). Such procedures are potentially useful in longitudinal social surveys for reconciling flows and cross-sectional estimates. Such accounting procedures have rarely been applied in social survey contexts, but Statistics Netherlands has used them to produce a consolidated picture of employment changes (Mushkudiani *et al*. 2015).

Detailed transaction data between businesses, measured by tax systems, are beginning to be analysed as networks to support national accounts, eg Dhyne *et al*. (2015). If longitudinal surveys can be linked to similar administrative data with the consent of respondents, then these analyses may also be extended to social surveys.

**Discussion**

As we show above there are many ways in which business survey data demonstrate longitudinal elements. Some standard approaches in business surveys make use of this information across time, but there are many opportunities for borrowing more structured longitudinal thinking from the methods developed for cohort and panel studies in the social sciences. Commonly used sampling processes for business surveys ensure that the largest businesses are completely enumerated (and therefore appear continuously in repeating surveys) and also create large overlaps in samples from period to period through rotation and sample coordination. These two conditions create a longitudinal structure. Although this is not generally the primary purpose of these surveys, it forms the basis for longitudinal methodology to be used. Composite estimation has been under-used in business surveys because of the requirement (perceived or real) for consistency between estimates of level and change, which have been met by periodic benchmarking and/or time series processes.

There are also many similarities in the procedures in business and social surveys for measuring changes; the stability of the questionnaire and data collection processes are important because they hold components of the non-sampling errors approximately constant and lead to unbiased estimates of changes. Both types of surveys have to deal with changes in structures (of person within households, and units within businesses), and have different ways to do this – longitudinal household panels with following rules and weight share methods, and business surveys by defining cross-sectional structures. There is scope here to extend the estimation for changes in longitudinally matched business datasets to use weight share methods, and even to design surveys or datasets to follow individual business units through changes in ownership and analyse the effects of conditions in their parent businesses.

The rather different conditions and characteristics demonstrated by business populations, such as their skewness and the availability of a sampling frame mean that the methods will need some adjustment, and this suggests avenues for future research, several of which we have highlighted. In some situations, the frame makes things easier – it gives an indication of eligibility for nonrespondents, for example, even in situations where the frame information may be out of date. Published research on longitudinal data appear to be largely restricted to social surveys – for example, we found no reports of attempts to measure rotation group bias or seam bias in business surveys, although several authors mention them in passing as if they are established phenomena.

Databases of longitudinally linked business data are becoming commoner, and this opens up a wide range of possibilities for more detailed longitudinal analysis. The terminology for estimating the time to transitions varies between subject areas, but there is scope for these methods to be applied in these datasets and for them to generate new insights.

We hope that this paper will encourage further exchange between practitioners of business and longitudinal surveys, and that this will widen the range of tools available to both groups and lead to some novel approaches and applications.

**Conflict of interest statement**

The Authors declare that there is no conflict of interest.

**Acknowledgements**

We are grateful to Boris Lorenc for encouraging us to pursue this topic, and to the two anonymous referees and Peter Lynn whose comments led to a significantly improved paper.